\documentclass[twocolumn,aps,prb,superscriptaddress,showpacs,floatfix,amsmath,amssymb
]{revtex4-2}

\usepackage{graphicx}
\usepackage{dcolumn}
\usepackage{bm}
\usepackage{color}
\usepackage{amsmath}
\usepackage{hyperref}

\begin{document}

\preprint{APS/123-QED}

\title{Current-Induced Modulation of Spin-Wave Propagation in a Y-Junction via Transverse Spin-Transfer Torque}

\author{Lorenzo Gnoatto}
\affiliation{Department of Applied Physics and Science Education, Eindhoven University of Technology, P.O. BOX 5132, 5600 MB Eindhoven, The Netherlands\\}

\author{Rai M. Menezes}%
\affiliation{COMMIT, Department of Physics, University of Antwerp, Groenenborgerlaan 171, B-2020 Antwerp, Belgium\\}
\affiliation{Departamento de Física, Universidade Federal de Pernambuco, Cidade Universitária, 50670-901 Recife-PE, Brazil\\}

\author{Artim L. Bassant}
\affiliation{Institute for Theoretical Physics, Utrecht University, Princetonplein 5, 3584 CC Utrecht, The Netherlands\\}

\author{Rembert A. Duine}
\affiliation{Department of Applied Physics and Science Education, Eindhoven University of Technology, P.O. BOX 5132, 5600 MB Eindhoven, The Netherlands\\}
\affiliation{Institute for Theoretical Physics, Utrecht University, Princetonplein 5, 3584 CC Utrecht, The Netherlands\\}

\author{Milorad V. Milo\v{s}evi\'c}%
\email{milorad.milosevic@uantwerpen.be}
\affiliation{COMMIT, Department of Physics, University of Antwerp, Groenenborgerlaan 171, B-2020 Antwerp, Belgium\\}

\author{Reinoud Lavrijsen}
\email{r.lavrijsen@tue.nl}
\affiliation{Department of Applied Physics and Science Education, Eindhoven University of Technology, P.O. BOX 5132, 5600 MB Eindhoven, The Netherlands\\}

\date{\today}

\begin{abstract}
We report the transverse control of spin-wave propagation in the configuration where the spin-wave wavevector \(\mathbf{k}\) is perpendicular to the charge-current density \(\mathbf{J}\). Building on theoretical predictions of spin-wave refraction by nonuniform spin-polarized currents, and guided by micromagnetic simulations used to optimize the device geometry and current distribution, we experimentally explore a Y-shaped Permalloy structure in which a locally injected current perturbs the spin-wave dispersion. Measurements reveal current-dependent amplitude differences between the two output branches, providing initial experimental indications consistent with transverse, spin-transfer-torque–driven deflection. Although the magnitude of the effect is modest and accompanied by significant uncertainties, the observed trends qualitatively follow expectations from the simulations. These results demonstrate the feasibility of influencing spin-wave routing through local current injection and establish a proof-of-concept basis for current-controlled manipulation of spin-wave propagation in reconfigurable magnonic circuits.
\end{abstract}

\maketitle


The ability to dynamically control SW propagation pathways is a critical milestone toward realizing reconfigurable magnonic circuits and advancing energy-efficient computing architectures~\cite{mahmoud2020introduction,Chumak2022Roadmap,chumak_magnon_2015}. A spin-wave multiplexer enables this functionality by selectively routing SW signals into different output channels based on an external control parameter, forming a fundamental building block for more complex devices such as magnonic logic gates~\cite{MZT_Interferometer}.

State of the art demonstrations include multiplexers based on active Oersted field manipulation of spin waves~\cite{vogt2014realization}, as well as directional couplers where signal routing is governed by frequency, power, and applied magnetic fields~\cite{wang2020magnonic,heussner2018frequency,heussner2020experimental}. 

In this work, we investigate a multiplexer concept based on Spin Transfer Torque (STT), in which a spin-polarized current locally perturbs the spin-wave dispersion and trajectory, offering a potential route toward active signal steering. This idea builds on our previous simulation studies, which indicated that spatially non-uniform spin-polarized current distributions can influence spin-wave propagation and may enable functionalities such as neuromorphic computing and multichannel routing~\cite{PhysRevApplied.22.054056}. Here, we examine how the current density distribution can act as an effective diffracting medium for spin waves: analogous to Snell’s law in optics \cite{SNELL_BACK}, gradients in the spin-polarized current and variations in the incidence angle are expected to produce distinct refraction-like deflections. This mechanism underpins the concept of current-controlled spin-wave steering. In the following sections, we present experimental data that exhibit signatures consistent with this behavior, while also discussing the limitations and uncertainties inherent to the presented measurements. In the present experimental implementation, direct spatial mapping of spin-wave refraction is not accessible; instead, we use the relative transmitted amplitudes at two outputs as an indirect observable sensitive to current-induced spin-wave routing.

\begin{figure}[t]
	\centering
	\includegraphics[width=1\linewidth]{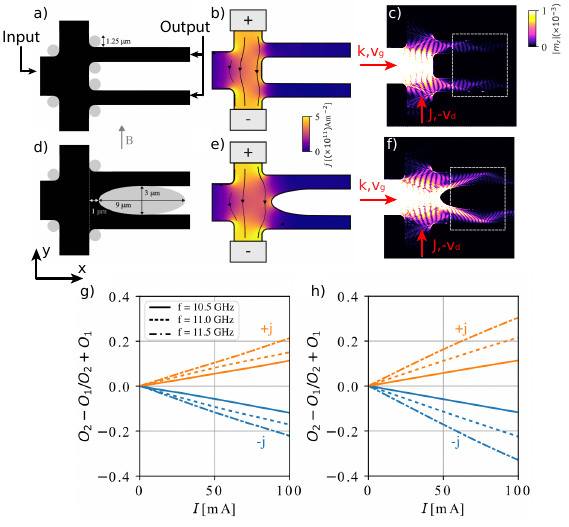}
	\caption{\label{SWD_1} Simulation setup illustrating the two device geometries, along with the corresponding current density, magnetization distribution, and normalized spin-wave output asymmetries. a) Schematic of the simulated geometry~g1 with rounded corners, showing the input/output ports and relevant geometric parameters.  
	b) Simulated current density distribution for geometry~g1.  
	c) Spin-wave magnetization amplitude across the constriction in geometry~g1, with the directions of wavevector \( \mathbf{k} \) and current density \( \mathbf{J} \) indicated. Here, the highlighted rectangle indicates the region where output transmissions are calculated.  
	d–f) Corresponding panels for geometry~g2.  
	g) Normalized output asymmetry \( (O_2 - O_1)/(O_2 + O_1) \) as a function of applied current for geometry~g1, evaluated at excitation frequencies of 10.5, 11.0, and 11.5~GHz.  
	h) Same as g), but for geometry~g2.} 
\end{figure}

To design and optimize the device geometry, we performed micromagnetic simulations using the $\mathrm{MuMax^3}$ framework. These simulations incorporate realistic current-density profiles, the resulting STT-induced torques, and their influence on spin-wave trajectories within bifurcated Y-shaped waveguides. Permalloy was chosen as the active material due to its comparatively high spin polarization, low magnetic damping and low electrical resistivity, which-relative to other common ferromagnets such as CoFeB-make it well suited for enhancing the STT-driven modulation of spin-wave propagation~\cite{gnoatto_2024}. In the device geometry shown in Figure~\ref{SWD_1}~(a-c), the spin polarized drift current $v_d$, carries angular momentum proportional to the current density~\cite{vlaminck2008current} and induces a spin-wave group velocity component $v_g$, along the current direction. The resulting routing of the spin-wave trajectory is proportional to the ratio $v_d/v_g$. Therefore, materials with high group velocity exhibit reduced sensitivity to current-induced routing for the same applied current. In our devices, at an applied magnetic field of 60\,mT, we measure a group velocity of $\sim$2.4 km/s for Py, consistent with values reported for Damon–Eshbach spin waves in similar Permalloy structures from time-resolved imaging studies~\cite{wessels_direct_2016}.

This study builds on the geometry proposed in Ref.~\cite{PhysRevApplied.22.054056}, where additional simulation details can be found. We reiterated two specific designs, shown in Fig.~\ref{SWD_1}a–c, to address several key physical constraints relevant for experimental implementation. The simulations were performed for a 20\,nm-thick Permalloy film using the following material parameters: saturation magnetization $M_{\mathrm{s}} = 0.75,\mathrm{MA/m}$, exchange stiffness $A = 10,\mathrm{pJ/m}$, gyromagnetic ratio $\gamma/2\pi = 29,\mathrm{GHz/T}$, and damping constant $\alpha = 7 \times 10^{-3}$~\cite{mahmoud2020introduction}. Dipolar interactions and Oersted fields were included, yielding a spin-wave dispersion for $k$-vectors in the range 1–6,$\mu\mathrm{m}^{-1}$, comparable to the values accessible in our experimental approach. 

\begin{figure}[t]
	\centering
	\includegraphics[width=\linewidth]{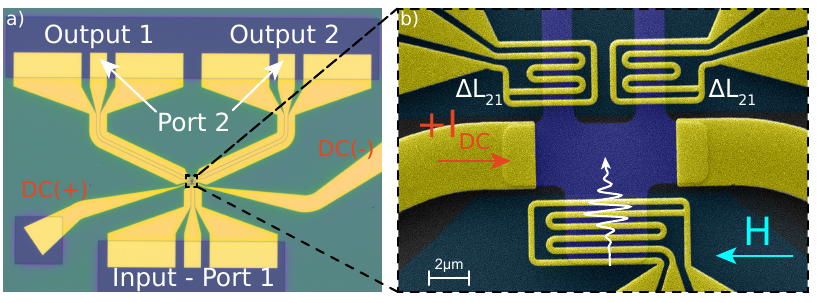}
	\caption{\label{SWD_2}a) Optical micrograph of the spin-wave multiplexer device, showing the input connected to VNA port 1 and the two outputs (\(O_1\)) and (\(O_2\)) connected to port 2. DC current contacts for positive and negative polarity are also labeled. 
		b) False-colored scanning electron micrograph of the central region. The direction of spin-wave propagation (white arrow), applied DC current (red arrow), and external magnetic field $\mu_0 \mathrm{H}$ (cyan arrow) are indicated. The two output branches correspond to transmitted signals $\Delta \mathrm{L_{21}}$, measured at different output antennas.} 
\end{figure}

To optimize the efficiency of spin-wave routing via spin-transfer torque, we focused on maximizing the current-induced spin wave amplitude $O$ output asymmetry $\frac{(O_2-O_1)}{(O_1 + O_2)}$, as shown in Fig.~\ref{SWD_1}g and h, between the two output branches ($O_1$ and $O_2$) as a function of current magnitude and polarity. First, for the Permalloy strip used here, we expect a spin-wave propagation length of approximately $(5\!-\!10~\mu\text{m})$ \cite{wessels_direct_2016}. To ensure reliable signal detection, the separation between the input and output antennas was chosen to match this length scale. We designed the device with an input-output antenna spacing (center to center) of $(D = 7~\mu\text{m}$, which provides an optimal compromise between maintaining a detectable spin-wave amplitude and ensuring a sufficient interaction length with the transverse spin-polarized current, aiming to maximize the effectiveness of the spin-transfer-torque modulation within spin-wave propagation limits. Second, the spatial separation and width of the output channels were selected to balance competing constraints. The two branches were kept as close as possible near the junction to ensure that most of the spin-wave energy emitted from the input could be collected, while still maintaining sufficient separation to avoid electrical contact between the antennas (see Fig.~\ref{SWD_2}b). Each output waveguide was designed with a width of \(1.5~\mu\text{m}\), representing a compromise informed by fabrication limits and detection sensitivity. Narrower widths (below \(1~\mu\text{m}\)) have not been successfully measured in our setup and could result in insufficient magnetic volume beneath the antenna for effective inductive detection, particularly in the presence of temperature increases from Joule heating, where \(M_{\mathrm{s}}\) may be reduced~\cite{gladii2016spin}. The interaction region width, i.e. the region where the spin-waves interact with transverse current, was fixed at \(4~\mu\text{m}\).

Based on these design constraints, we compared two device geometries with the dual objective of maximizing the transmitted spin-wave signal in the absence of current and enhancing the current-induced output amplitude asymmetry between the two branches. The first configuration (g1), shown in Fig.~\ref{SWD_1}a–c, features rounded corners at the waveguide junction to reduce current crowding and mitigate the risk of electromigration, which typically arises at sharp corners due to localized enhancements in current density. The corresponding distributions of spin-polarized current density and spin-wave amplitude are shown in Fig.~\ref{SWD_1}b–c, with the detection region indicated by the highlighted rectangle.

The second geometry (g2), shown in Fig.~\ref{SWD_1}d–f, was designed with smoother curvature at the junction to promote improved spin-wave transmission by minimizing reflections at the bifurcation. Micromagnetic simulations confirm that geometry~g2 indeed yields a higher transmitted amplitude and stronger current-induced asymmetry (Fig.~\ref{SWD_1}g–h), indicating more efficient spin-wave routing. However, the simulations also reveal that the continuous curvature in g2 introduces a slight tilt of the propagating spin-wave fronts. This loss of wavefront coherence becomes detrimental in the antenna detection region, where optimal inductive coupling requires the spin-wave vector to remain perpendicular to the experimental meandered antenna lines. As a result, despite its superior transmission, geometry~g2 is less compatible with our detection scheme requiring straight wave-fronts at the location of the pick-up antenna's.

For these reasons, and to avoid experimentally problematic tilts of the spin-wave front, we proceeded with the fabrication of devices based on geometry~g1, as described in the next section. We note that the current-induced asymmetry increases at higher frequencies, consistent with the reduction in spin-wave group velocity. 

\begin{figure}[t]
	\centering
	\includegraphics[width=\linewidth]{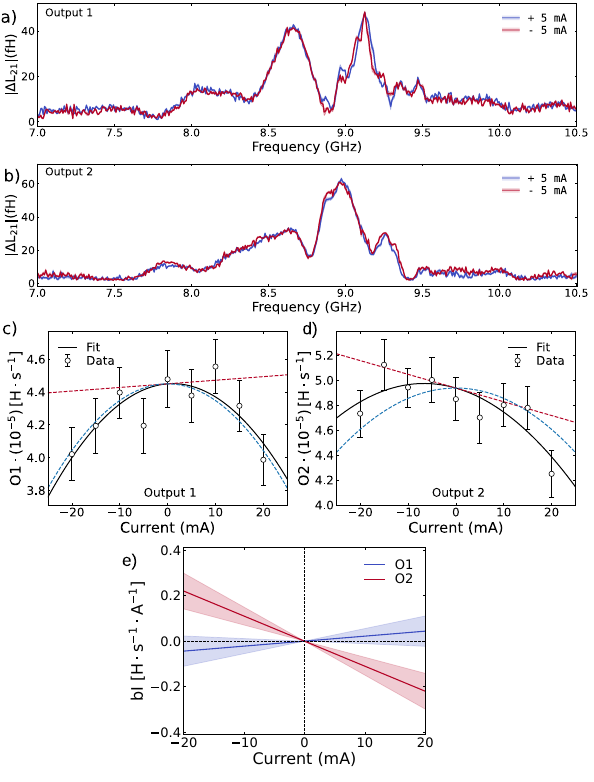}
	\caption{\label{SWD_3} a–b)  Transmitted signal amplitude for \(O_1\) and \(O_2\), respectively. c–d) Integrated signal area for \(O_1\) and \(O_2\) as a function of applied current, with fits showing linear (red) and quadratic (blue) components. e) Extracted linear component for both \(O_1\) and \(O_2\), with corresponding error bands.} 
\end{figure}

Devices were fabricated following the procedures outlined in Ref.~\cite{gnoatto_2024}, using a ferromagnetic stack consisting of Ta(4\,nm)/Py(20\,nm)/Ta(4\,nm). A wide-field optical micrograph of a representative device is shown in Fig.~\ref{SWD_2}a. The input waveguide is connected to port~1 of a Vector Network Analyser (VNA), while the two output arms, Output~1 (\(O_1\)) and Output~2 (\(O_2\)), can be connected to port~2, interchangeably. The DC contact pads used for current injection are also indicated. Outputs \(O_1\) and \(O_2\) cannot be measured simultaneously, as our setup cannot accommodate more than two antenna probes at once; each output was therefore characterized independently. The spin-wave group velocity decreases with increasing applied magnetic field. Although a lower group velocity enhances the interaction time between the spin waves and the spin-polarized current—thereby strengthening the spin-transfer torque (STT) effect—it also reduces the propagation length, i.e. the amplitdue of signal at the detector, limiting the measurable amplitude. To balance these competing effects, we selected the highest magnetic field that still provided sufficient spin-wave transmission. This optimization led to the choice of a uniform external field of \(\mu_{0}H = 60~\mathrm{mT}\) and an excitation frequency range of 7–10.5~GHz for all measurements. Representative transmission spectra for \(O_1\) and \(O_2\), recorded under positive and negative DC currents (\(\pm 5~\mathrm{mA}\)), are shown in Fig.~\ref{SWD_3}a–b. Each spectrum was averaged seven times, with the VNA internally averaging 20 times per point. Standard errors in the extracted amplitude values are indicated and defined later in the text.

To ensure device stability and rule out time-dependent effects such as thermal drift or current-induced training, each sample was first subjected to a preconditioning cycle. This step mitigates potential property changes arising from Joule heating or sustained high current densities—such as electromigration at local hot spots—which are relevant concerns given the operating conditions of our devices. Following the procedure described in Ref.~\cite{gnoatto_2024}, we applied \(+25~\mathrm{mA}\) (corresponding to a maximum current density of \(J = 4.95 \times 10^{11}~\mathrm{A/m^{2}}\)) for 8~hours, followed by \(-25~\mathrm{mA}\) for an additional 8~hours. This protocol also served as a stress test to confirm the device’s robustness under prolonged operation. In Section~I of the Supplemental Material~\cite{SM}, we present the device response at zero current after each current cycle, demonstrating that the properties remain stable throughout the experiment.

Notably, the mode profiles measured at outputs \(O_1\) and \(O_2\) (see Fig.~\ref{SWD_3}a–b) differ significantly from each other and from the nearly Gaussian-like lineshapes commonly observed in straight magnonic waveguides~\cite{chumak_magnon_2015}. This deviation is likely a direct consequence of the Y-shaped junction geometry, which introduces strong spatial variations in the internal demagnetizing field. As spin waves propagate through the bifurcation, the resulting inhomogeneous field landscape can excite additional modes—such as edge-guided or transverse standing waves—that distort the spectral response and reduce overall coherence. Small angular misalignments of the applied magnetic field or minor fabrication imperfections may further modify the local demagnetizing environment at the junction, contributing to the observed spectral asymmetries. A full modal decomposition would be required to quantify these contributions, but is beyond the scope of this study.

To enable a direct comparison between simulations and experimental results, we quantify the output asymmetry using the experimental ratio \(\frac{O_2 - O_1}{O_1 + O_2}\), where the values of \(O_1\) and \(O_2\) are obtained by integrating the area under the respective transmission spectra. To first order, this integrated amplitude is proportional to the number of magnons transmitted from the input to each output antenna, and can be used as a measure of the net flux of spin angular momentum carried by the spin waves into \(O_1\) and \(O_2\).

For each data point, the standard error of the mean is computed as \(\mathrm{SEM} = \sigma / \sqrt{N}\), where \(\sigma\) is the standard deviation of the measured amplitudes and \(N = 7\) is the number of repetitions. The uncertainty in the integrated area is estimated by calculating upper and lower bounds using the modified integrands,$ A_{\mathrm{upper}} = \int (A(f) + \mathrm{SEM}) \, \mathrm{d}f, \qquad
A_{\mathrm{lower}} = \int (A(f) - \mathrm{SEM}) \, \mathrm{d}f,$ and taking \(\delta A = (A_{\mathrm{upper}} - A_{\mathrm{lower}})/2\). The resulting averaged areas and their uncertainty intervals are plotted in Fig.~\ref{SWD_3}a–b.

This procedure provides an estimate of the total transmitted spin-wave intensity and allows us to assess whether the applied current breaks the amplitude symmetry between the two outputs, as predicted by the simulations in Fig.~\ref{SWD_2}. Based on the transverse STT routing mechanism, we expect the linear current dependence to exhibit opposite slopes for \(O_1\) and \(O_2\), corresponding to preferential routing of spin waves toward one branch depending on the current polarity.

Two main contributions are expected in the current-dependent spin-wave transmission: one arising from Joule heating and the other from spin-transfer torque (STT). Joule heating depends on the magnitude of the applied current and results in symmetric changes in signal amplitude for positive and negative polarities (as discussed in Ref.~\cite{gladii2016spin}). In contrast, the STT-induced routing effect depends on the current direction, and is therefore expected to introduce an antisymmetric contribution to the transmitted signal. 

To separate these effects, the integrated spin-wave signals were fitted using the expression $A(I) = aI^2 + bI + c$, where the quadratic term ($aI^2$) captures the Joule heating contribution and the linear term ($bI$) captures the STT-driven asymmetry. Figure~\ref{SWD_3}c–d displays the resulting fits, with the quadratic and linear components highlighted in red and blue, respectively.

The extracted linear terms are shown with uncertainties in Figure~\ref{SWD_3}e. As anticipated, \(O_1\) and \(O_2\) exhibit linear components of opposite sign, consistent with current-induced routing.  Although the magnitude of the linear term is modest and carries sizeable uncertainty, the opposite-sign trends of \(O_1\) and \(O_2\) are consistent with the expected signature of transverse-STT-induced routing. Additional devices exhibiting similar behavior are shown in section II of Supplemental Material~\cite{SM}.

In conclusion, this work explores an approach for current-controlled rerouting of spin-wave propagation in a Y-shaped geometry using transverse spin-transfer torque. Guided by micromagnetic simulations incorporating realistic current-density distributions, we identified device parameters that enhance the detectability of spin-wave signals and the sensitivity to current-induced modulation. In the fabricated structures, we observe measurable current-dependent differences in the amplitudes detected at the two output channels. These results provide initial experimental indications of transverse spin-wave routing driven by spin-polarized currents, consistent with the qualitative trends predicted by simulations.

Although the magnitude of the effect is modest and accompanied by significant uncertainties, the observations demonstrate the feasibility of influencing spin-wave routing through local current injection. This establishes a proof-of-concept foundation for future efforts toward selective activation of spin-wave pathways, which may ultimately contribute to reconfigurable magnonic signal-processing and logic architectures. The quantitative uncertainty and limited number of devices measured prevent a definitive assessment of the routing strength, but the qualitative trends merit further investigation. Further improvements in device design e.g. that would symmetrize the outputs at zero current, measurement statistics, reversed field experiments, mode control and mode understanding will be essential to fully quantify and harness this mechanism.

\textit{Acknowledgments}. This work was supported by the Dutch Research Council
(NWO) under "OCENW.KLEIN.502 Black Holes on a chip". This work was supported by the Research Foundation—Flanders (FWO-Vlaanderen), EoS ShapeME project, INCT project Advanced Quantum Materials, involving the Brazilian agencies CNPq (Proc. 408766/2024-7), FAPESP, and CAPES.

\bibliography{apssamp}


\end{document}


\preprint{APS/123-QED}

\title{Supplemental Material:\\ Current-Induced Modulation of Spin-Wave Propagation in a Y-Junction via Transverse Spin-Transfer Torque}

\author{Lorenzo Gnoatto}
\affiliation{Department of Applied Physics and Science Education, Eindhoven University of Technology, P.O. BOX 5132, 5600 MB Eindhoven, The Netherlands\\}

\author{Rai M. Menezes}%
\affiliation{COMMIT, Department of Physics, University of Antwerp, Groenenborgerlaan 171, B-2020 Antwerp, Belgium\\}
\affiliation{Departamento de Física, Universidade Federal de Pernambuco, Cidade Universitária, 50670-901 Recife-PE, Brazil\\}

\author{Artim L. Bassant}
\affiliation{Institute for Theoretical Physics, Utrecht University, Princetonplein 5, 3584 CC Utrecht, The Netherlands\\}

\author{Rembert A. Duine}
\affiliation{Department of Applied Physics and Science Education, Eindhoven University of Technology, P.O. BOX 5132, 5600 MB Eindhoven, The Netherlands\\}
\affiliation{Institute for Theoretical Physics, Utrecht University, Princetonplein 5, 3584 CC Utrecht, The Netherlands\\}

\author{Milorad V. Milo\v{s}evi\'c}%
\email{milorad.milosevic@uantwerpen.be}
\affiliation{COMMIT, Department of Physics, University of Antwerp, Groenenborgerlaan 171, B-2020 Antwerp, Belgium\\}

\author{Reinoud Lavrijsen}
\email{r.lavrijsen@tue.nl}
\affiliation{Department of Applied Physics and Science Education, Eindhoven University of Technology, P.O. BOX 5132, 5600 MB Eindhoven, The Netherlands\\}

\date{\today}
\maketitle

\section{Area variation during measurements}\label{app:area_var}
\begin{figure}[ht]
	\centering
	\includegraphics[width=0.5\linewidth]{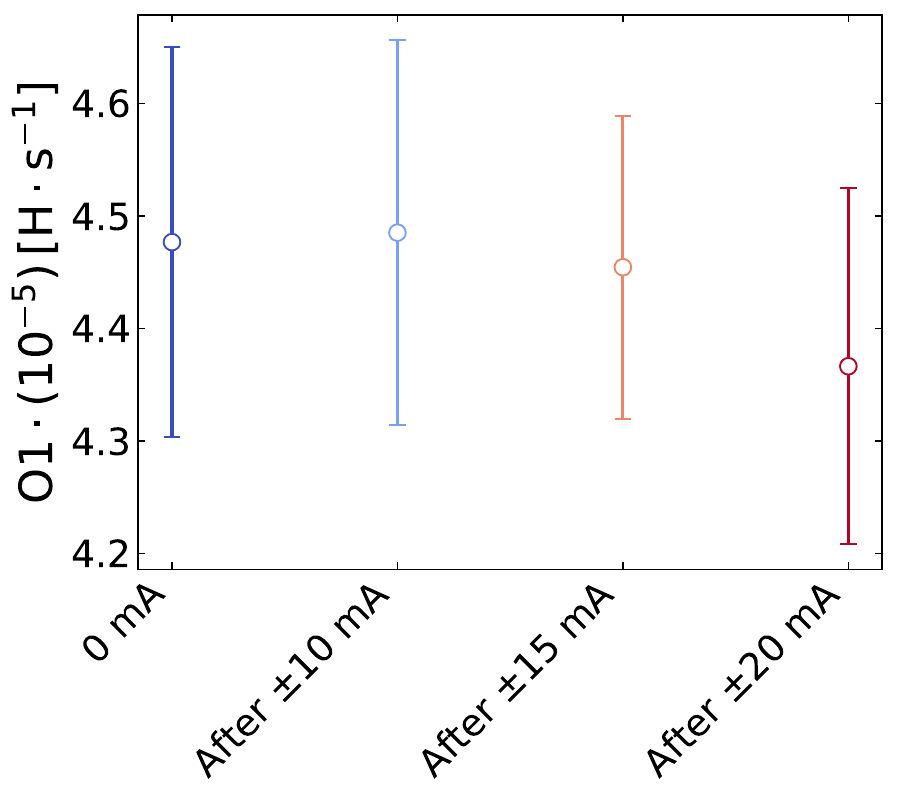}
	\caption{\label{SWD_sup_2} Measurements at 0 mA after a full set of measurements ($\pm 10\text{mA},~\pm 15\text{mA},~\pm 20\text{mA})$ was performed for $O_1$ of the device measured in Figure 3 (c-e) in the main text. } 
\end{figure}

To verify device integrity and assess reproducibility throughout the experiment, we monitored the rerouted spin-wave signal at zero DC current following each set of measurements performed under increasing current amplitudes. The measurement sequence followed the order: 0~mA, $\pm$5~mA, 0~mA, $\pm$10~mA, 0~mA, and so on. This allowed us to track any potential degradation or irreversible changes in the signal due to current stress. As shown in Fig.~\ref{SWD_sup_2}, the extracted signal areas at zero current remain largely consistent over the measurement sequence. While a slight downward trend in the mean values is observed, the overlap of the standard error of the mean (SEM) intervals suggests that this variation is not statistically significant within the current dataset.

\section{Additional measurements on different devices}\label{app:add_meas}
\begin{figure}[ht]
	\centering
	\includegraphics[width=\linewidth]{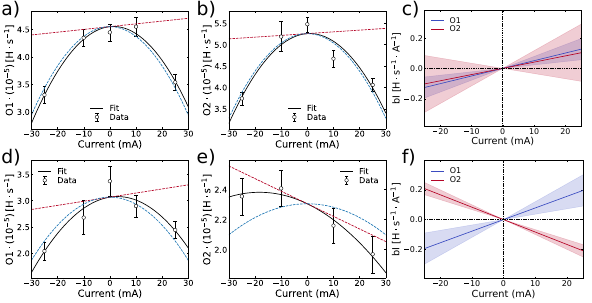}
	\caption{\label{SWD_sup_1}  a–b) Transmitted signal amplitude for $O_1$ and $O_2$, respectively, for a second measured device; c) corresponding linear component comparison. d–f) Same analysis for a third device.} 
\end{figure}

Figures~\ref{SWD_sup_1}a–c and~\ref{SWD_sup_1}d–f present measurements from two additional devices, used to assess the reproducibility of the spin-wave rerouting behavior. While both datasets qualitatively show the expected superposition of a symmetric quadratic background (attributed to Joule heating) and a polarity-dependent linear contribution (linked to spin-transfer torque), the strength of the linear response varies between devices. In particular, the device shown in Figure~\ref{SWD_sup_1}a–c exhibits significantly larger uncertainties in the extracted linear term, making it difficult to confidently resolve the directionality of the STT-induced rerouting. In contrast, the device in Figure~\ref{SWD_sup_1}d–f shows a clearer and more consistent trend, similar to that reported in the main text. The origin of the larger uncertainty in device a–c is difficult to pinpoint, as all devices were measured using identical protocols and equipment. Nevertheless, the overall behavior supports the conclusion that transverse STT enables rerouting of spin-waves.

